\documentclass[aps,pra,groupedaddress, 
twocolumn,nofootinbib,superscriptaddress]{revtex4-1}
\newcommand{\be}{\begin{eqnarray}}
\newcommand{\ee}{\end{eqnarray}}
\newcommand{\beq}{\begin{equation}}
\newcommand{\eeq}{\end{equation}}

\usepackage{graphicx,epsfig}
\usepackage{amsmath}
\usepackage{amsfonts}
\usepackage{fancyhdr}
\usepackage{amsmath}
\usepackage{mathrsfs}
\usepackage[utf8]{inputenc} % Accents i altres símbols

\newcommand{\Ket}[1]{\left| #1 \right>}\newcommand{\bra}[1]{\left< #1 \right|}
\newcommand{\braket}[2]{\left< #1 \big |  #2\right>}

\usepackage{color}
\begin{document}

  %\preprint{}

\title{Tunneling vortex dynamics in linearly coupled Bose-Hubbard rings}

\author{Albert Escriv\`a}
\email{albert.escriva@fqa.ub.edu}
\affiliation{Departament de F\'isica Qu\`antica i Astrof\'isica, 
Facultat de F\'isica, Universitat de Barcelona, 
Mart\'i i Franqu\`es 1, 08028 Barcelona, Spain}
\affiliation{Institut de Ci\`encies del Cosmos, Universitat de Barcelona, 
Mart\'i i Franqu\`es 1, 08028 Barcelona, Spain}

\author{Antonio Mu\~noz Mateo}
\affiliation{Departament de F\'isica Qu\`antica i Astrof\'isica, 
Facultat de F\'isica, Universitat de Barcelona, 
Mart\'i i Franqu\`es 1, 08028 Barcelona, Spain}
\affiliation{Graduate School, China Academy of Engineering Physics,
100193 Beijing, China}

\author{Montserrat Guilleumas}
\affiliation{Departament de F\'isica Qu\`antica i Astrof\'isica, 
Facultat de F\'isica, Universitat de Barcelona, 
Mart\'i i Franqu\`es 1, 08028 Barcelona, Spain}
\affiliation{Institut de Ci\`encies del Cosmos, Universitat de Barcelona, 
Mart\'i i Franqu\`es 1, 08028 Barcelona, Spain}

\author{Bruno Juli\'a-D\'iaz}
\affiliation{Departament de F\'isica Qu\`antica i Astrof\'isica, 
Facultat de F\'isica, Universitat de Barcelona, 
Mart\'i i Franqu\`es 1, 08028 Barcelona, Spain}
\affiliation{Institut de Ci\`encies del Cosmos, Universitat de Barcelona, 
Mart\'i i Franqu\`es 1, 08028 Barcelona, Spain}
% 
%\date{\today}
\begin{abstract}
The quantum dynamics of population-balanced fractional vortices and 
population-imbalanced vortices in an effective two-state bosonic 
system, made of two coupled discrete circuits with few sites, is 
addressed within the Bose-Hubbard model. 
We show that for low on-site interaction, the tunneling of quantized 
vortices between the rings performs a coherent, oscillating dynamics 
connecting current states with chiral symmetry. The vortex-flux 
transfer dually follows the usual sinusoidal particle current of 
the Josephson effect, in good agreement with a mean-field 
approximation. Within such regime, the switch of persistent currents in the 
rings resembles flux-qubit features, and is feasible to experimental 
realization. On the contrary, strong interatomic interactions suppress the 
chiral current and lead the system into fragmented condensation.
\end{abstract}
%\pacs{}
%
\maketitle
\section{Introduction}

The recent advances in the preparation and control of ultracold 
atomic gases have fostered the appearance of a specific subfield 
termed atomtronics, which is aimed at the development of 
technological applications based on atomic matter 
waves~\cite{atomtronics,atomtronics2,atomtronicexperimental1}. 
Its first steps are mainly pursuing successfully-applied 
theories from the fields of quantum optics and electronics, but 
also exploiting the ability of degenerate quantum gases to 
simulate other complex quantum systems~\cite{quantumsimulation}.

In the search of basic tools that could eventually realize 
functional atomtronic devices, superconducting technologies are 
specially inspiring. This is due to the fact that the underlying 
condensates of superconducting electron pairs find their counterpart 
in the electrically neutral, ultracold Bose-Einstein condensates (BECs) 
of bosonic atoms, or even of paired fermionic atoms in the BCS-BEC 
crossover. Since the first experimental realizations of ring-trapped 
BECs, atomic circuits can be built to support persistent 
currents~\cite{Ryu2007,Ramanathan2011,Moulder2012,Beattie2013}, and 
also to mimic the performance of the versatile superconducting 
quantum interference devices (SQUIDs) in atomtronic 
devices~\cite{Eckel2014,Ryu2013}. In particular, the feasibility of 
trapping ultracold atoms in looped geometries allows the bosonic 
systems to explore the emulation of the superconducting flux qubits 
used in quantum computing~\cite{Chiorescu2003}. 
Flux qubits are based on the switch of persistent currents 
flowing through Josephson junctions in a looped circuit threaded 
by magnetic flux. The role of the magnetic fluxoid in superconducting 
devices can be played by rotation-induced vortices in neutral 
bosonic systems. The entry or the exit of a vortex in the loop 
induces a change in the quantized circulating current that leads 
the system into a different metastable state. This scenario can 
lead to a candidate flux qubit if the two states connected by the 
vortex transit are sufficiently isolated, and if a coherent quantum 
dynamics of the system, oscillating between the two states, is 
achievable.

The search for plausible realizations of bosonic flux qubits 
has attracted great attention in the ultracold atomic gas 
community. Several proposals have explored looped one-dimensional 
geometries both in 
discrete~\cite{amico1,Aghamalyan2015,Hallwood2011,gallemi1}
and continuous systems~\cite{Schenke2011,Halkyard2010}.
Within a mean-field framework, Refs.~\cite{amico1, amico3} have 
elaborated on plausible implementations of bosonic qubits. In 
spinor condensates, Refs.~\cite{gallemi3,Tylutki2016,Calderaro2017}
showed the coherent 
transfer of half vortices between the condensate components, 
Refs.~\cite{lesanovsky,BRAND1} analyzed Josephson oscillations 
in the angular momentum, whereas in Ref.~\cite{andrea}
the dynamics of quantized currents has also been studied 
within the Bogoliubov approximation.
On the other hand, there are many works 
based on a Bose-Hubbard (BH) ladder, where vortices can be 
generated by means of artificial magnetic 
fields~\cite{hugel2014,tokuno,piraud,uchino,NICOLAS}. The system 
dynamics can evolve into the one dimensional equivalent of a 
superconducting vortex lattice~\cite{orignac} with an associated 
Meissner-vortex phase transition~\cite{amico2}. Recently, this 
ultracold-gas Meissner effect has been realized in lab 
experiments~\cite{chiral_antonio}.

The present work contributes to the search of coherent quantum 
dynamics of vortices in bosonic circuits. Our setting makes use 
of two linearly coupled, discrete rings with a few number of 
sites, whose static properties have been previously reported 
in Ref.~\cite{ours1}.  By means of a BH model, we explore the 
quantum dynamics of population-imbalanced vortices and 
population-balanced fractional vortices in double rings with different 
intra- to inter-ring coupling ratios and on-site interaction 
strengths. Different from regular, stationary vortex states, we 
search for realistic parameters in non-equilibrium, interacting 
systems of this type that allow for a regime of coherent oscillations 
of the initial vortex phase between the two rings. We explore 
different dynamical regimes and we show that there exists a 
range of parameters for low interactions where the transfer of 
vortex states between the two rings is coherent.

The paper is organized as follows: In Sect.~\ref{sec2} we describe 
the Hamiltonian of the system and present its characteristic parameters; 
we also introduce the analytical vortex solutions of the 
single-particle problem that will be used to build many-particle 
imbalanced vortices and fractional vortices. In Sect.~\ref{sec3} we address 
the many-body problem of such non-stationary vortex solutions and 
identify their different dynamical regimes. Finally, Sect.~\ref{sec4} 
gathers our conclusions and prospections for future work.

\section{Theoretical model}
\label{sec2}

We consider $N$ bosons loaded in two parallel Bose-Hubbard rings 
with $M$ sites per ring. The coupling  between rings, $J_\perp$, 
connects only sites with equal azimuthal coordinate. Inside each 
ring, there is a coupling $J$ between next-neighbor sites. As a 
result, the system is described by the following Hamiltonian,
\be
\label{hamiltonian}
 \mathcal{\hat{H}} \, &&=  \sum_{l=0}^{M-1} \left[\, -J\,
\sum_{j=\uparrow,\downarrow} (\hat{a}^{\dagger}_{l,j} \, 
\hat{a}_{l+1,j}+\hat{a}^{\dagger}_{l+1,j} \,\hat{a}_{l,j} ) 
\right. \\
 && \left. - J_{\perp} 
\,(\hat{a}^{\dagger}_{l,\uparrow}\,\hat{a}_{l,\downarrow} 
+ \hat{a}^{\dagger}_{l,\downarrow}\,\hat{a}_{l,\uparrow}) +
 \frac{U}{2} 
 \sum_{j=\uparrow,\downarrow}\hat{n}_{l,j}(\hat{n}_{l ,j}-1) \right] \,,
\nonumber
\ee
where the bosonic creation and annihilation operators for the site 
$l$ in the ring $j$ (with $j=\uparrow,\downarrow$) are 
$\hat{a}^{\dagger}_{l,j}$ and $\hat{a}_{l,j}$, respectively. They 
fulfill the canonical commutation relations 
$[\hat{a}_{l,j},\hat{a}^{\dagger}_{m,k}]=\delta_{lm}\delta_{jk}$. 
The corresponding number operators are 
$\hat{n}_{l,j}=\hat{a}^{\dagger}_{l,j} \hat{a}_{l,j}$, and the on-site 
atom-atom contact interaction strength is assumed to be repulsive $U>0$.

From the Hamiltonian ~(\ref{hamiltonian}), we construct the time 
evolution operator, $\hat{U}(t)= \exp{(-i \hat{\cal H} t/\hbar)}$, 
that propagates in real time an initial many-body state with $N$ 
bosons, $\Ket{\Psi(t=0)}$, as 
$\Ket{\Psi(t)} = \hat{U}(t) \Ket{\Psi(t=0)}$. We compute this action 
by means of the SciPy implementation of the algorithm developed in 
Ref.~\cite{algoritmo}. At fixed 
time during the dynamical evolution, we measure the population 
imbalance between rings $z(t)$, the transition amplitude to a 
particular state $P(t)$, and the chiral current $\hat{L}_{\rm chi}(t)$, 
which informs us on the angular momentum (vorticity) imbalance between rings. 

The population imbalance $z(t) \in [-1,1]$ is calculated as, 
\beq
z(t)={N_{\uparrow}(t)-N_{\downarrow}(t) \over N} \,,
\eeq
where $N_{j}(t)=\sum_{l=0}^{M-1}\bra{\Psi(t)} {\hat n}_{l,j}\Ket{\Psi(t)}$ 
is the average number of atoms in the $j$ ring. The total number of 
atoms in the system is a conserved quantity, 
$N=N_{\uparrow}(t)+N_{\downarrow}(t)$.

The probability of finding a particular target state, 
$\Ket{\Psi_{\rm target}}$ reads, 
\beq
P(t) = |\braket{\Psi_{\rm target}}{\Psi(t)}|^2\,.
\eeq 
The total azimuthal current is given by 
$\hat{L}= \hat{L}_{\uparrow}+\hat{L}_{\downarrow}$,  where the 
azimuthal current in ring $j$ is given by
\beq
\hat{L}_{j} = -i \,{J \over \hbar} \sum_{l=0}^{M-1} 
(\hat{a}^{\dagger}_{l,j} 
\hat{a}_{l+1,j} - 
\hat{a}^{\dagger}_{l+1,j} \hat{a}_{l,j})\,.
\eeq
From these operators, we define the chiral current operator as 
the difference in azimuthal current (or relative azimuthal current) 
between the two rings 
$\hat{L}_{\rm chi}= \hat{L}_{\uparrow}-\hat{L}_{\downarrow}$, and 
compute the mean chiral current, normalized to its initial value, as 
the non-dimensional quantity
\beq
L_{\rm chi}(t)={\bra{\Psi(t)} \hat{L}_{\rm chi} 
\Ket{\Psi(t)} \over  |\bra{\Psi(0)} \hat{L}_{\rm chi} \Ket{\Psi(0)}|} \,.
\eeq

The condensed fraction and the fragmentation of an $N$-particle 
many-body state is characterized by the normalized eigenvalues $p_l=N_{l}/N$, 
where $N_l$ are the eigenvalues of the one-body 
density matrix $\hat{\rho}$~\cite{condensed_fractions}. The matrix 
elements of the latter are given by
\begin{equation}
\rho_{(l,j),(m,k)}=\bra{\Psi} {\hat a}^{\dagger}_{l,j} 
\, {\hat a}_{m,k} \Ket{\Psi} \,,
\label{one-body}
\end{equation}
so that 
%${\rm Tr} (\hat \rho) =$
$\sum_{l=1}^{2M} p_l=1$.  We will henceforth refer to fragmentation when
there are more than one eigenstates of order one, $p_l\propto \mathcal{O}(1)$, even though the small number of particles in the systems considered does not properly allows us to state that
there exists a macroscopic occupation of eigenstates.

\subsubsection{Single-particle vortices and fractional vortices} 

The single-particle dispersion (at $U=0$) of the Hamiltonian 
(\ref{hamiltonian}) contains two energy branches 
$\epsilon_q^\pm=-2J\cos(2\pi\,q/M)\mp J_\perp$ that correspond to Bloch 
waves~\cite{hugel2014, ours1,Gil2019}
\begin{equation}
\Ket{\Psi_{q}^{\pm}}=
\frac{1}{\sqrt{2M}}\sum_{l=0}^{M-1}
e^{\textstyle{i\frac{\,2\pi q \, l}{M}}} 
\left(\hat{a}_{l,\uparrow}^{\dagger}\pm \hat{a}_{l,\downarrow}^{\dagger}\right) 
\Ket{\rm vac} \,,
\label{Bloch}
\end{equation}
where the integer quasimomentum takes the values
$q=0,\,\pm 1,\,\pm 2,\dots,\lfloor M/2\rfloor$, e.g. for $M=3$, $q=0, \pm 1$ 
and for $M=4$, $q=0,\pm 1, 2$. 
Both energy branches $\epsilon_q^\pm$ belong to 
eigenstates of the total current operator with eigenvalues 
$L_q= 2J\sin(2\pi\,q/M)/\hbar$ 
and present zero chiral current.
For large number of sites $M$, the mentioned eigenvalues tend to 
$L_{q}/ (4 \pi\,J/M\hbar)=q$. We will refer to these states  
as stationary (or regular) vortices of charge $q$. The vortices 
$\Ket{\Psi_{q}^{-}}$ in the higher energy branch $\epsilon_q^-$
present $\pi$-phase shifted rings. 

When the inter-ring coupling tends to zero $J_{\perp}\rightarrow 0$, 
the eigenfunctions $\Ket{\Psi_{q}^\pm}$ tend to be energetically 
degenerate, and the linear combinations  
$(\Ket{\Psi_{q}^{+}}\pm\Ket{\Psi_{q}^{-}})/\sqrt{2}$ 
approximate stationary states with currents localized in one of the 
rings $j$, explicitly,
\begin{equation}
\Ket{\Psi_{q,j}} = \hat\Psi^\dagger_{q,j} \Ket{\mathrm{vac}} \equiv   
\frac{1}{\sqrt{M}} \sum_{l=0}^{M-1} 
\textstyle{e^{\textstyle{i\frac{2\pi q  \,l}{M}}}} 
\hat{a}^{\dagger}_{l,j}\Ket{\rm vac} \,,
\label{initial_states0}
\end{equation}
whereas the other ring is empty. These states are only stationary 
in the (non-interacting) limit of decoupled rings, and will be 
referred as vortices of charge $q$ with full imbalance $z=1$ ($z=-1$), 
when $j=\uparrow$ ($\downarrow$). 

We also consider states that combine two different single vortices of the 
type (\ref{initial_states0}), with charges $q$ and $q'$, which are 
loaded one in each ring to give a balanced $z=0$ system as
\begin{eqnarray}
&&\Ket{\Psi_{(q,q')}}=
\Ket{\Psi_{q,\uparrow}}\otimes \Ket{\Psi_{q'\!,\downarrow}}= \nonumber \\
&&\left( \sum_{l=0}^{M-1} 
\frac{ \textstyle{e^{\textstyle{i\frac{2\pi q  \,l}{M}}}} }{\sqrt{M}}
\hat{a}^{\dagger}_{l,\uparrow}\right)
\otimes\left( \sum_{l=0}^{M-1} 
\frac{ \textstyle{e^{\textstyle{i\frac{2\pi q'  \,l}{M}}}}  }{\sqrt{M}}
\hat{a}^{\dagger}_{l,\downarrow} \right)\, 
\Ket{\rm vac} \,.
\label{eq:qq1}
\end{eqnarray}
Again, the resulting states are stationary only in the limit of 
(non-interacting) decoupled rings. The particular case 
with $q=1$ and $q'=0$ is the minimal example, carrying half the current of a 
regular vortex $L_q/2$, and corresponds to the
so-called half quantum vortex of a continuous system
(see e.g~\cite{Tylutki2016,Calderaro2017}, and~\cite{Son2002} for
the properties of these vortex states, and their
relation with domain walls of the relative phase in two-component
condensates).
For general $(q,q')$, due to the fractional value of the associated total 
azimuthal current, we will refer to states~(\ref{eq:qq1}) 
as population balanced, fractional-vortex states. Note that these states, as 
well as the imbalanced vortices discussed before, carry in the general case 
both nonzero total azimuthal current (as regular vortices of Eq.~(\ref{Bloch})) 
and nonzero chiral current (different from regular vortices).

\subsubsection{Mean-field ansatz for the large-particle-number limit}

The effect of the population imbalance on the many-body dynamics, as present in 
the imbalanced vortex states of Eq~(\ref{initial_states0}), 
can be characterized within a mean-field approximation when the total 
number of particles $N$ is large. In this case, the many-body state 
can be expressed as a coherent macroscopic superposition of equal 
states in the top, 
$\Ket{\Psi_{q,\uparrow}}=\hat\Psi^\dagger_{q,\uparrow}\Ket{\rm vac}$, 
and in the bottom, 
$\Ket{\Psi_{q,\downarrow}}=\hat\Psi^\dagger_{q,\downarrow}\Ket{\rm vac}$, 
rings~\cite{trialwave}:
\begin{align}
\Ket{\Psi (t)} =&\frac{1}{\sqrt{N!}}\left(\sqrt{\frac{1+z}{2}}\, 
e^{i \phi/2} \, \hat\Psi^\dagger_{q,\uparrow}\right. \,\nonumber \\
&+\left.\sqrt{\frac{1-z}{2}}\,e^{-i \phi / 2} 
\,\hat\Psi^\dagger_{q,\downarrow}\right)^{N}\Ket{\rm vac} \,,
\label{mf-state}
\end{align}
where $z(t)$ is the population imbalance and 
$\phi(t)$ is the average phase difference between the two rings. 
From this ansatz, the expectation value of the Hamiltonian~(\ref{hamiltonian})
$\langle {\cal H} \rangle={\bra{\Psi} \hat{\cal H} \Ket{\Psi}}/{NJ_{\perp}}$, in units of $NJ_{\perp}$, can be approximated by
\begin{equation}
\langle {\cal H} \rangle= 
\frac{\epsilon_q }{J_{\perp}}-\sqrt{1-z^2}\,\cos \phi 
+\frac{M\Lambda}{2}(1+z^2) \,,
\label{semi}
\end{equation}
where $\epsilon_q=-2J\cos(2\pi\,q/M)$. The imbalance and the relative phase are canonically conjugate variables.
From Eq.~(\ref{semi}) the following equations of motion can be obtained:
\begin{equation}
\begin{aligned}
\frac{d z}{d \tilde{t}}&= -\sqrt{1-z^2}\,\sin \phi\,,\\
\frac{d \phi}{d \tilde{t}} &= M\Lambda z +\frac{z}{\sqrt{1-z^2}}\,\cos \phi \,,
\label{eq:BJJ}
\end{aligned}
\end{equation}
where we have defined the dimensionless time $\tilde{t}$, in units 
of the Rabi period $t_R=\pi\hbar/J_\perp$, and the dimensionless 
interaction parameter $\Lambda$ by
\begin{align}
\tilde{t} &= \frac{2\pi t}{t_R} = \frac{2J_{\perp}}{\hbar}\,t  \, ,\\
\Lambda &= \frac{(N-1)}{2\,M}\frac{U}{J_\perp} \,.
\label{eq:param}
\end{align}
Note that (\ref{eq:BJJ}) are the typical equations of a single, short 
bosonic Josephson junction~\cite{amico1}, since the ansatz 
(\ref{mf-state}) involves only average quantities of the two 
rings. In spite of the fact that the real dynamics of the 
system corresponds instead to a long Josephson 
junction~\cite{Gil2019} (where the tunneling between rings is 
site dependent), we will show that this mean-field ansatz provides a 
good approximation when the interparticle interactions are low 
(against tunneling) and the evolution is constrained to a few 
periods of $t_R$. In any case, as we will see, the natural 
parameters $t_R$ and $\Lambda$ of this model are relevant parameters 
for identifying the dynamical regimes of the system. 

\section{Imbalanced-vortex and fractional-vortex dynamics}
\label{sec3}
\begin{figure}[t]
\includegraphics[width=1\linewidth]{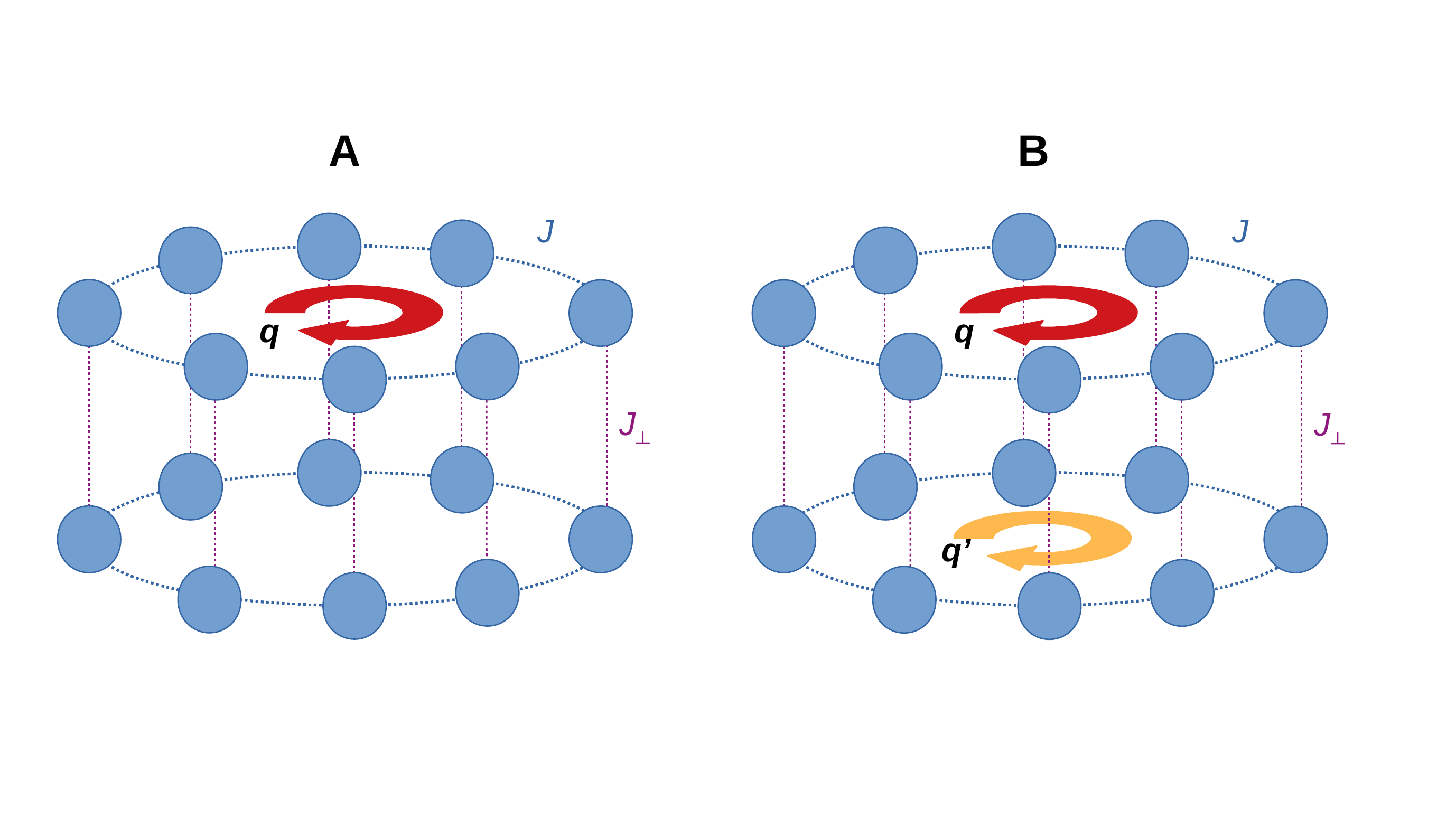} 
\caption{Schematic picture of the two different initial states 
considered. Configuration A (left): fully population-imbalanced 
($z=1$) vortex state of charge $q$ in the top ring. Configuration B 
(right): population-balanced fractional vortex of charge $q$ and $q'$ 
in the top and bottom rings, respectively.
\label{4rings}}
\end{figure}

We investigate the coherent quantum tunneling between the 
two coupled rings of population-imbalanced vortices and 
population-balanced fractional vortices. We consider as initial 
configuration the two types of many-body states 
schematically depicted in Fig.~\ref{4rings}. In configuration A 
(on the left of Fig.~\ref{4rings}), we prepare a fully imbalanced, 
$z(0)=1$, vortex state of charge $q$ in the top ring, whereas the 
bottom ring is initially unpopulated. On the other hand, in 
configuration B (on the right of Fig.~\ref{4rings}), we prepare a 
fully balanced state, $z(0)=0$, that combines different vortices of charges 
$q$ and $q'$ in different rings. In both configurations, unless explicitly stated
otherwise, we will assume that the  preparation 
of the initial state takes place at zero inter-ring coupling $J_\perp=0$, 
and zero interaction strength $U=0$. The subsequent time evolution of 
the system is monitored after switching on, instantaneously, particular 
non-zero values of $J_\perp$ and $U$. We search for the regime of 
coherent exchange of phase between the two coupled rings. 

For completeness,
we will also present two cases that simulate a more realistic experimental
procedure by using as initial states interacting stationary vortices, so that
only a single parameter has to be instantaneously switched on. As  we will see,
maybe counter-intuitively, these alternative initial conditions make a marginal
difference with respect to the non-interacting ansatz in the observed dynamics,
reflecting that the key feature for the few-particle systems considered is encoded
 in the translational invariance and phase profile of the initial vortex state. 

For later comparison, it is useful to analyze first the tunneling dynamics 
of single-particle vortices. Let us assume an initial configuration 
of type A, without population in the bottom ring:
\begin{equation}
\Ket{\Psi^{(q)}(t=0)} =\Ket{\Psi_{q , \uparrow}} \,.
\label{eq:single_v}
\end{equation}
After switching on a non-zero coupling $J_\perp$, the time evolution 
of this state shows the coherent transfer of vorticity between the 
rings according to
\begin{eqnarray}
\Ket{\Psi^{(q)}(t)}=e^{\textstyle{-i\frac{\epsilon_q t}{\hbar}}} \, &&
\left[\, \cos(J_{\perp}t / \hbar)\,\Ket{\Psi_{q,\uparrow}} \right. + \nonumber \\
i\,&&\left. \sin(J_{\perp}t/\hbar)\,\Ket{\Psi_{q,\downarrow}} \,\right]\,.
\label{time_state_single}
\end{eqnarray}
The population imbalance evolves periodically in time 
\be
z(t) = \cos({2 J_\perp \, t}/{\hbar}) \,,
\label{eq:popA}
\ee
and the probability amplitude to find a fully imbalanced vortex 
in the bottom ring can be obtained from the target state 
$\Ket{\Psi_{\rm target}}=\Ket{\Psi_{q,\downarrow}}$ as 
\beq
P(t) = \left| \braket{\Psi_{\rm target}}{\Psi^{(q)}(t)}\right|^2= \sin^2( 
J_{\perp}t/\hbar)\,.
\label{eq:transA}
\eeq 
The non-normalized chiral current, that accounts for the 
phase (hence momentum) exchange between rings, reads
\begin{equation}
L_{\rm chi}^{(q)}(t) =\frac{2J}{\hbar} \, \sin\left(\frac{2\pi q}{M}\right) \,
\cos({2 J_\perp t}/{\hbar})\,,
\label{eq:chiA} 
\end{equation}
and follows the population exchange (\ref{eq:popA}). For large number 
of sites, $2\pi q/M\ll 1$, the initial chiral current tends to 
$\bra{\Psi_{q , \uparrow}}\hat{L}_{\rm chi} 
\Ket{\Psi_{q , \uparrow}} =(4\pi J/\hbar M) q$, which gives the relative azimuthal 
current between the two rings, and is quantized in integer units of 
$(4\pi J/\hbar M)$.

As expected, this is the typical dynamical evolution of two linearly 
coupled quantum systems. The single-particle results are independent 
of the number of sites, and (when normalized to initial state values) 
also of the intra-ring tunneling rate $J$ and vortex charge $q$. 
We will see that this coherent scenario cannot always be kept 
when interparticle interactions are taken into account.

It is also interesting to compare this single particle estimate with the semiclassical
prediction (\ref{eq:BJJ}). For vanishing interaction parameter $\Lambda\rightarrow 0$, 
the mean-field time evolution of the imbalance is harmonic
$\ddot z + \omega^2 z=0$, with angular frequency $\omega=2\pi t_R^{-1}$, 
hence consistent with (\ref{eq:popA}). The opposite limit of dominant interaction
leads to anharmonic oscillations $\ddot z+ z\,\sqrt{1-z^2} M\Lambda\cos\phi=0$ that
at high imbalance $z\rightarrow 1$  are suppressed ($\ddot z\approx \dot z \approx 0$) in a self-trapping regime.

For configuration B, the initial single-particle fractional vortices 
\begin{equation}
\Ket{\Psi^{(q,q')}(t=0)} 
= \Ket{\Psi_{q,\uparrow}}\otimes \Ket{\Psi_{q',\downarrow}} 
 \,,
\label{initial_1B}
\end{equation}
evolve as
\begin{align}
\Ket{\Psi^{(q,q')}(t) \!}=
e^{\textstyle{-i\frac{\epsilon_q t}{\hbar}}}\left[\cos(J_{\perp}t / \hbar)\Ket{\Psi_{q,\uparrow}}  +
i\sin(J_{\perp}t/\hbar)\Ket{\Psi_{q,\downarrow}}\right] \nonumber\\
\otimes \; e^{\textstyle{-i\frac{\epsilon_{q'} t}{\hbar}}} \left[ \cos(J_{\perp}t/\hbar)
\Ket{\Psi_{q',\downarrow}}+i\sin(J_{\perp}t/\hbar)\Ket{\Psi_{q',\uparrow}}
\right ] \,,
\label{2v-t}
\end{align}
and the population balance, $z=0$, persists during the whole time 
evolution. From the target state $|{\Psi}_{\rm target}\rangle = 
\hat\Psi^\dagger_{q',\uparrow}\otimes\hat\Psi^\dagger_{q,\downarrow} \Ket{\rm vac}$,
the resulting transition probability, which monitors the transfer of the
vortices, produces the same result (\ref{eq:transA}) as configuration A. 
And the same happens for the non-normalized chiral current
\begin{align}
 L_{\rm chi}^{(q,q')}(t) = \frac{2 J}{\hbar} \left[\sin\left(\frac{2 \pi 
q}{M}\right)-\sin\left(\frac{2 \pi q'}{M}\right)\right]\cos\left({2 J_\perp 
t}/{\hbar} \right)\,,
\label{eq:chiB}
\end{align}
when normalized to its initial value. Therefore, after normalization, 
the non-interacting phase-current dynamics turns out to be independent 
of the initial state choice A or B, the latter one irrespective of 
the selected values $q$ and $q'$, as well. Again, for large number of 
sites when $2\pi q/M\ll 1$ and $2\pi q'/M\ll 1$, Eq.~(\ref{eq:chiB}) 
tends at $t=0$ to $[ \bra{\Psi_{q',\downarrow}}\otimes \bra{\Psi_{q,\uparrow}}]
\hat{L}_{\rm chi} 
[\Ket{\Psi_{q,\uparrow}}\otimes \Ket{\Psi_{q',\downarrow}}]=(4\pi J/\hbar 
M)(q-q')$.
\begin{figure*}[t]
\includegraphics[width=\linewidth]{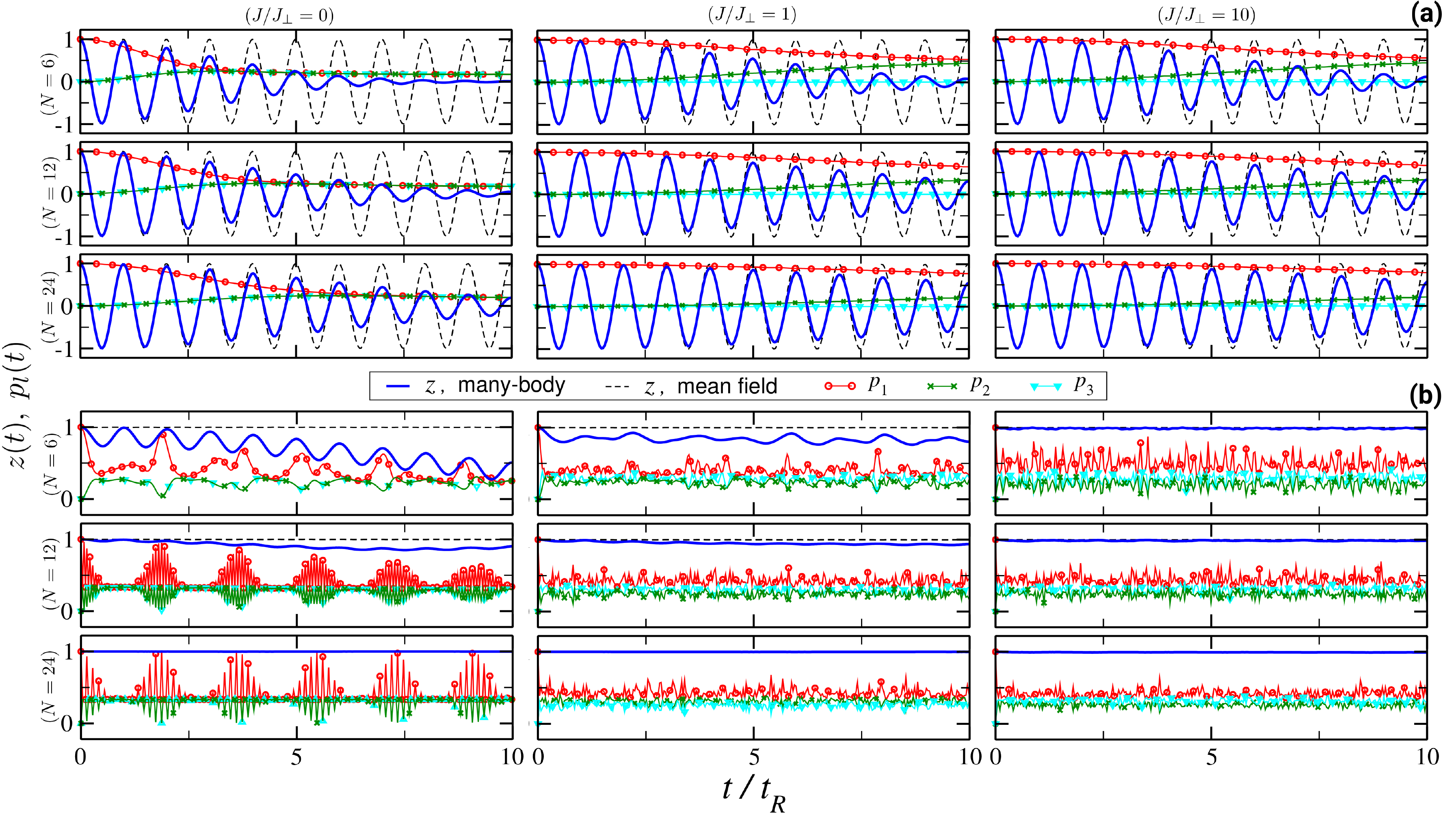}
\caption{Dynamical evolution of a fully imbalanced vortex of charge 
$q=1$ for different values of $J/J_{\perp}$, $N$ and $U$. Top panels (a) 
correspond to $\Lambda=0.2$, whereas bottom panels (b) correspond to
$\Lambda=50$. In all cases, the 
number of sites of each ring is $M=3$. The numerical solution for 
the population imbalance of the many-body quantum system 
(thick blue lines) is compared with the mean field approximation 
given by Eqs.~(\ref{eq:BJJ}) (thin black dashed lines). The three largest 
eigenvalues of the one-body density matrix $p_1$ (red circles), 
$p_2$ (green crosses), and $p_3$ (cyan triangles) are 
also plotted to monitor the condensed fraction and fragmentation of 
the system. The left, middle and right panel groups correspond to 
intra-ring couplings $J/J_{\perp}=0,\,1,\,10$, respectively, for 
inter-ring coupling fixed to $J_{\perp}=1$ in all panels. Time 
is measured in $t_R=\pi\hbar/J_\perp$ units.}
\label{time-evolution1}
\end{figure*}

\subsection{Many-body population-imbalanced vortex}
\label{situation_A}

We consider the initial many-body state 
to be a vortex of charge $q$ in the top ring with imbalance $z(0)=1$ 
(configuration A). That is, at $t=0$ all the atoms are populating the 
vortex state in the top ring and the bottom one is empty. In 
an $N$-particle bosonic system, in analogy with 
Eq.~(\ref{initial_states0}), the non-interacting $\Lambda=0$, 
population-imbalanced vortex is given by 
\begin{equation}
\Ket{{\Psi}_N^{(q)}}=\frac{1}{\sqrt{N!}} 
\left( \hat\Psi^\dagger_{q,\uparrow}  \right)^{N}\,\Ket{\rm vac}.
\label{eq:iniA}
\end{equation}
It is worth remarking that this is not an eigenstate of the total 
Hamiltonian Eq.~(\ref{hamiltonian}), even in the non-interacting 
case, since its evolution
\begin{align}
\Ket{\Psi_N^{(q)}(t)}_{\Lambda=0}=&\frac{1}{\sqrt{N!}}
\left(e^{\textstyle{-i\frac{\epsilon_q t}{\hbar}}}[\cos(J_{\perp}t/\hbar)\hat\Psi^\dagger_{q,\uparrow} \,+\right. 
\,\nonumber \\
&\left. 
i\sin(J_{\perp}t/\hbar)\hat\Psi^\dagger_{q,\downarrow}]\right)^{N}\Ket{\rm 
vac} \,,
\label{evolved_A}
\end{align}
replicates the single-particle coherent transfer of the vortex 
state between the two rings, with transition probability
to the many-body target state $\Ket{{\Psi}_{\rm 
target}}=({1}/{\sqrt{N!}})(\hat\Psi^\dagger_{q,\downarrow})^{N}\,\Ket{\rm vac}$ given 
by $P(t)=[\sin(J_{\perp}t/\hbar)]^{2N}$. 

Figure~\ref{time-evolution1} shows our numerical 
results for the time evolution of the many-body state, 
Eq.~(\ref{eq:iniA}), with vortex charge $q=1$, in a double ring 
system with $M=3$ and $J_\perp=1$, for two different 
interparticle-interaction values parameterized by 
$\Lambda=0.2$ and $\Lambda=50$. The population imbalance 
(thick solid lines), as well as the three largest eigenvalues of the 
one-body density matrix, $p_1, p_2$ and $p_3$ (symbols), are represented 
as a function of time. 
The left, middle and right panels correspond to different values 
of the intra-ring tunneling ratio $J/J_\perp= 0, 1$ and $10$, 
respectively. Inside each panel, thus for each interaction and 
coupling values, different number of particles $N=6, 12$ and $24$ 
(top, middle and bottom, respectively) have been considered. The 
mean-field limit value of the imbalance, obtained by solving the 
two-coupled equations (\ref{eq:BJJ}), is also shown (thin dashed lines) 
for comparison.
The latter features a bosonic Josephson junction within two distinct 
dynamical regimes determined by the values of 
$\Lambda$~\cite{junction}. For small interactions $\Lambda=0.2$ 
(top panels (a) of Fig.~\ref{time-evolution1}) the dynamical evolution
corresponds to a Josephson regime, where the population is 
coherently transferred between rings with oscillating imbalance 
around zero. On the contrary, for $\Lambda=50$ (bottom panels (b) of 
Fig.~\ref{time-evolution1}), the system enters the self-trapping 
regime, where the particles remain mostly localized in one of the 
two rings. As can be seen, the many-body dynamics approaches more 
to the mean field solution for increasing number of 
particles~\cite{PhysRevA.81.023615}. 

By construction, the initial state (\ref{eq:iniA}) is fully condensed 
($p_1=1,p_2=p_3=0$). However, after switching on the interaction, the 
system loses its coherence and becomes fragmented during the time 
evolution, which is manifested by the lost of the imbalance amplitude 
with respect to the sinusoidal mean-field value and $p_1 < 1$. The 
opposite situation is achieved within the Josephson regime in the 
limit case of $J/J_{\perp}=0$ (left panel in the top row of 
Fig.~\ref{time-evolution1}), where the system tends at long times to 
a fully fragmented situation ($p_{l}=1/6,\, \forall l$). The absence 
of intra-ring coupling $J=0$ leads to tri-fragmentation inside each 
ring at a first stage, after which the imbalance gets damped.

Within the Josephson regime, the presence of intra-ring coupling 
$J/J_{\perp} >0 $ (middle and right panels of the top row (a) in 
Fig.~\ref{time-evolution1}) stretches the duration of the coherence 
($p_1 \simeq 1$) exchange of phase and population between the rings. 
The resulting scenario does not change qualitatively with the 
particular value of nonzero $J$, and the long time tendency shows 
that the lost of coherence leads to a bifragmented state 
($p_1=p_2\simeq 1/2$ and $p_3=0$) reflecting the dynamics of two 
independent rings. 
%IS THIS SO?............
For longer times, not shown in the figure, revivals of the coherent 
oscillations are observed~\cite{revival,Milburn1997,Greiner2002}. We have checked that such 
revivals appear 
%sooner 
before for higher interaction values. 

The large interaction $\Lambda =50$ case, depicted in the bottom panel
row of Fig.~\ref{time-evolution1} (b), is generally well characterized by 
the mean-field prediction of population self-trapping, and only a 
slight departure from this behavior is observed for $J=0$ and $N=6$, 
with the lowest number of particles. As a consequence, the 
coherence dynamical regimes shown in the low interaction case are not 
reached at high interaction.

\begin{figure}[t]
\includegraphics[width=1\linewidth]{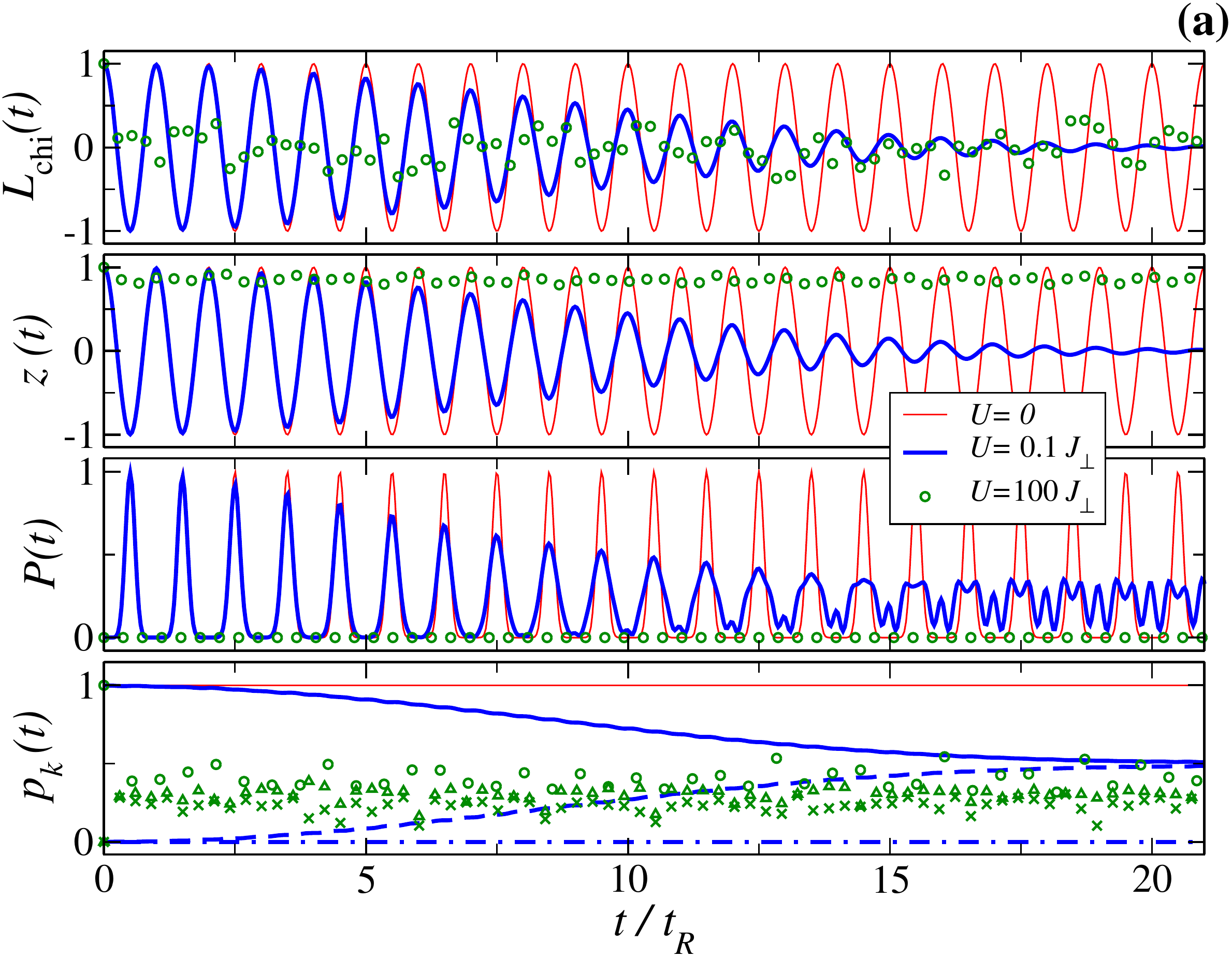} 
\includegraphics[width=\linewidth]{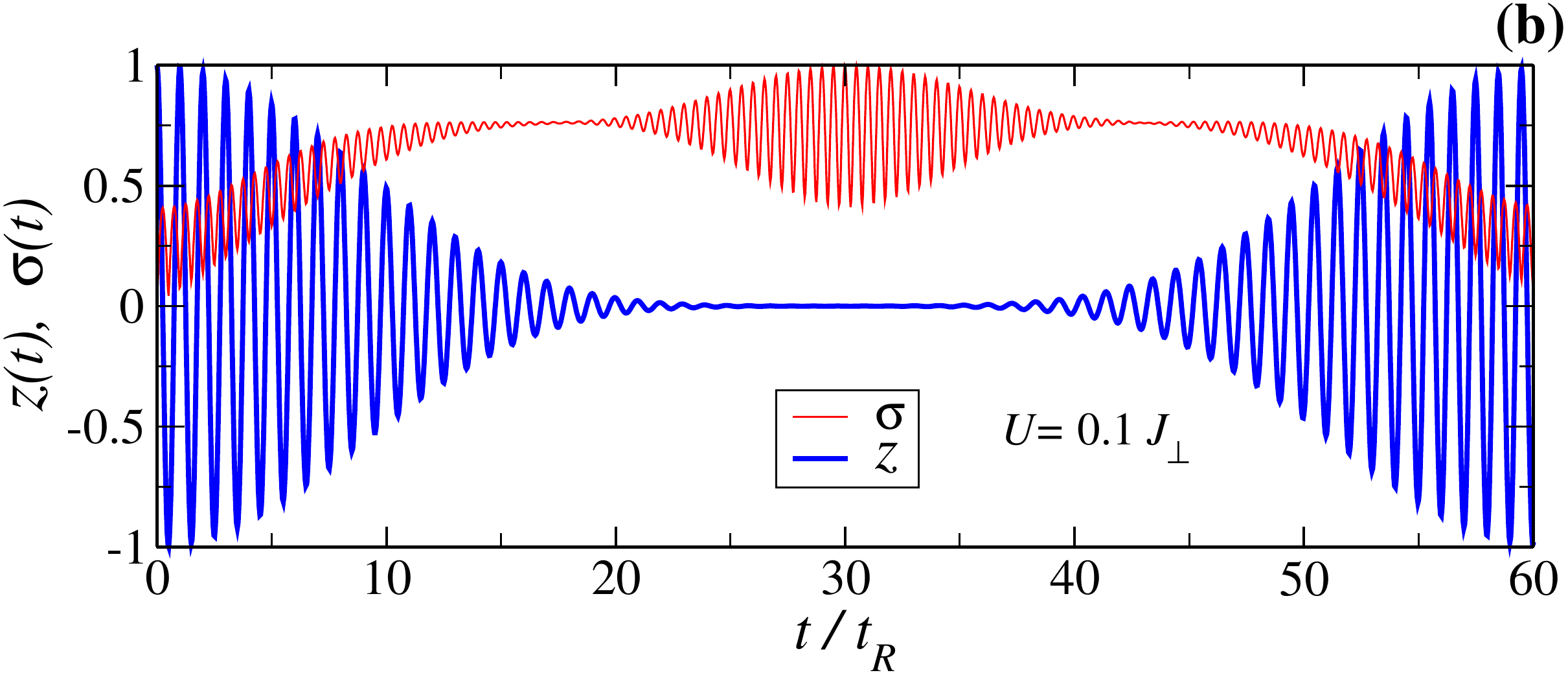} 
\caption{(a) Dynamical observables during the time evolution of a 
fully imbalanced vortex state with $q=1$, $N=6$, $M=3$, and 
$J/J_{\perp}=1$, for several values of the ratio 
$U/J_{\perp}$: $0$ (thin red line), $0.1$ (thick blue lines), 
$100$ (symbols). From top to bottom: Normalized 
chiral current, population imbalance, transition probability, 
and the three largest eigenvalues of the one-body density 
matrix ($p_{1}$, $p_2$, $p_3$). 
(b) Long time evolution of the
population imbalance, mean value
 (thick blue curve) and fluctuation (thin red curve), for the same setting and $U/J_{\perp}=0.1$,
showing the revival of the initial quantum state. Different from panel (a), the
initial state is an interacting stationary vortex in a single ring.
Time is measured in $t_R=\pi\hbar/J_\perp$ units.}
\label{vortex-groundA_1}
\end{figure}

Additional details of the effects of the interatomic interaction 
on the system dynamics are provided in Fig.~\ref{vortex-groundA_1}. 
The numerical time evolution of the chiral current, population 
imbalance, transition probability, and density-matrix eigenvalues, 
is represented for a system with $N=6$ atoms in the initial 
configuration A, with $q=1$ and $J/J_\perp=1$. Interaction values 
$U/J_\perp=0.1,$ (thick blue lines) and $100$ (symbols) have been chosen
 representing the Josephson and 
self-trapping regimes, and $U=0$ (thin red lines) is also depicted for comparison. 
Although in the presence of interaction, fragmentation cannot be 
avoided during the time evolution, the coherent transfer of vorticity 
is still possible for short times at low interaction, when the 
inter-tunneling coupling dominates the dynamics. The transition 
probability provides a good signature of the lack of coherence at 
long times even for low interaction. As can be seen in the figure, 
the characteristic sinusoidal character of the vortex transfer is 
lost around $t\simeq 10\,t_R$, when there is not a single dominant 
eigenvalue of the density matrix.

To obtain a more complete picture of the quantum dynamics,
Fig. \ref{vortex-groundA_1}(b) exhibits the average and fluctuation
values of the population imbalance for $U/J_{\perp}=0.1$,
during a long time evolution that includes the revival of the initial
 quantum state. Contrasting with panel (a), and for the sake of
 simulating more realistic experimental conditions, an interacting stationary
 vortex state in a single ring has been prepared as initial state. Although the outcome
 is almost indistinguishable from the evolution shown in panel (a), in this way
 only one parameter, the inter-ring coupling $J_\perp$, had to be 
 suddenly turned on at the beginning of the time evolution.
 As can be seen, the uncertainty in the average imbalance 
 $\sigma=\sqrt{\langle z^2\rangle-\langle z\rangle^2}$  grows when the 
 semi-classical approximation fails, at the end of the coherent oscillations.
Afterwards, interestingly, there is a regime of maximum variation in the
uncertainty associated with a
  zero average imbalance that precedes the revival of the initial quantum state,
 in agreement with the literature~\cite{revival,Greiner2002}. During the
 whole evolution, the uncertainty oscillates twice as fast as
 the average imbalance, reaching  minima when the absolute value of the average
 imbalance is maximum, and maxima when the average imbalance is zero.
\begin{figure}[t]
\includegraphics[width=1\linewidth]{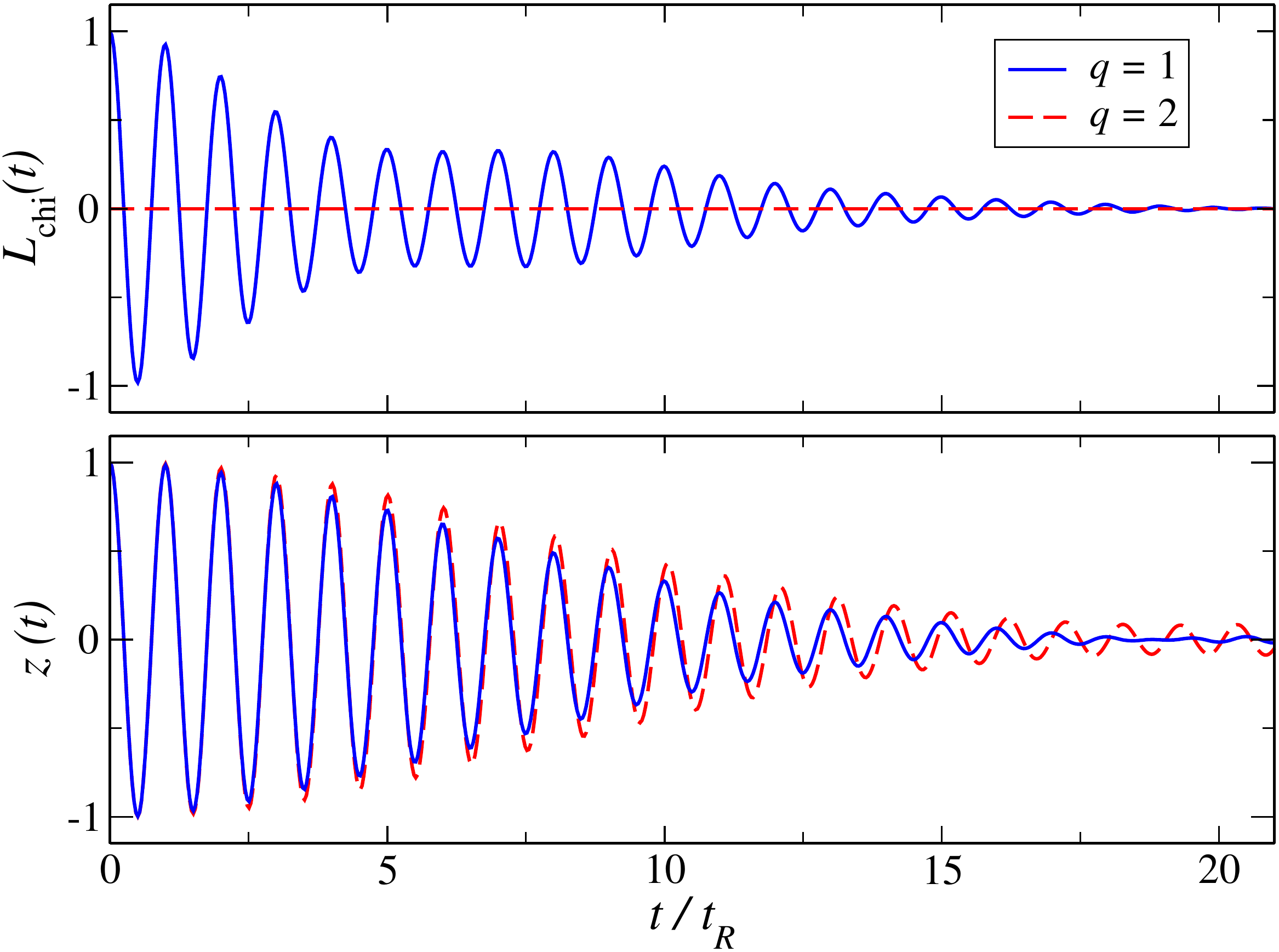} 
\caption{Dynamical evolution of the normalized chiral 
current $L_{\rm chi}(t)$ and the population imbalance $z(t)$ 
of a fully imbalanced vortex state of charge $q=1$ 
(blue solid line),  and $q=2$ (red dashed line) in a setting
with $U/J_{\perp}=0.1$, $J/J_{\perp}=1$, $N=6$, 
and  $M=4$.
}
\label{vortex-groundA_2}
\end{figure}

We have also considered the dynamics of a population-imbalanced 
vortex (initial configuration A)  in an non-commensurate system, 
having $M=4$ sites per ring and $N=6$ atoms,
within the Josephson regime of $U/J_\perp=0.1$ and $J/J_\perp=1$. 
This system allows one to compare between different initial 
vortex charges: $q=\pm1$ and $q=2$ 
(see Fig.~\ref{vortex-groundA_2}). Although the latter state does 
not carry any azimuthal current (since it lies at the edge of the 
Brillouin zone determined by the discrete lattice~\cite{Munoz2019}) 
and also $L_{\rm chi}(t)=0$, the time evolution of its population imbalance 
presents practically the same oscillatory behavior as the singly 
quantized vortices.

\subsection{Many-body population-balanced fractional vortices}
\label{situation_B}

We consider now the initial configuration B: the same number of 
atoms $N/2$ per ring, and different vortices of charge $q$ and $q'$ in the 
rings. The initial many-body state is described by
\begin{equation}
\Ket{{\Psi}_N^{(q,q')}}=\frac{1}{({N}/{2})!} \left[
\left( \hat\Psi^\dagger_{q,\uparrow}  \right)^{\frac{N}{2}}   \otimes
\left(\hat\Psi^\dagger_{q',\downarrow}\right)^{\frac{N}{2}} \right]
\,\Ket{\rm vac}\,.
\label{eq:iniB}
\end{equation}
This initial state is 
bifragmented with 
$p_1(t=0)=p_2(0)=1/2$.

\begin{figure}[t]
\includegraphics[width=1\linewidth]{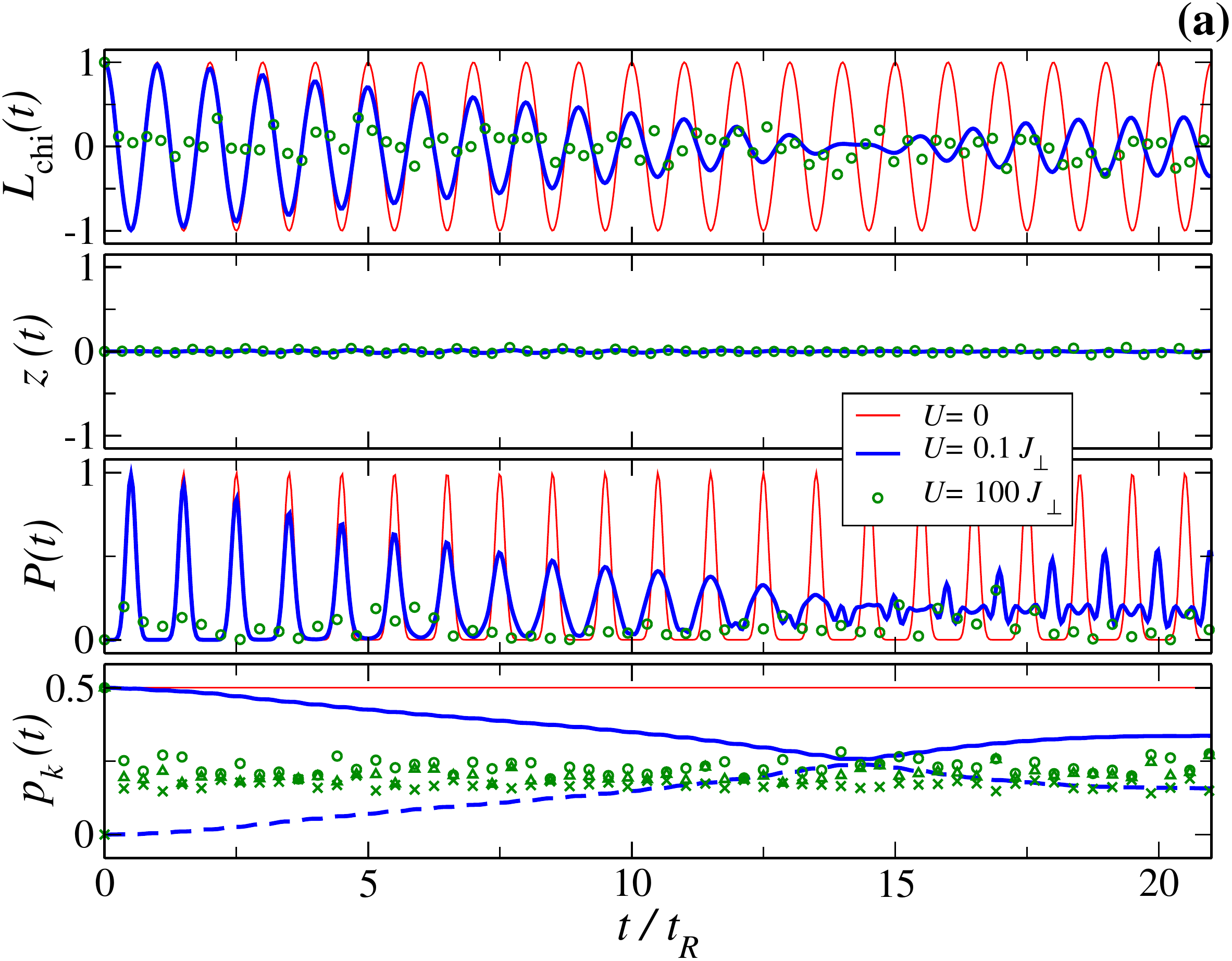} 
\includegraphics[width=1\linewidth]{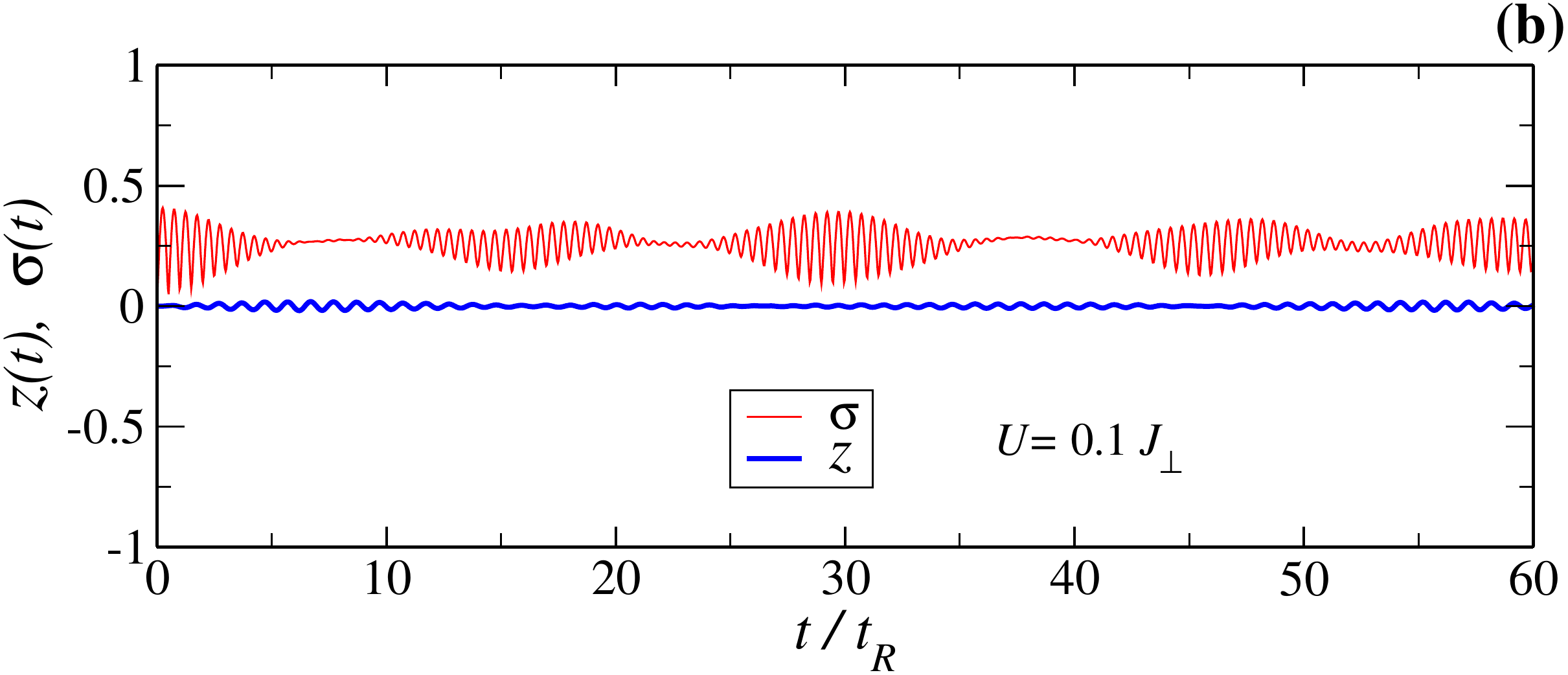} 
\caption{Same as Fig.~\ref{vortex-groundA_1}, in a setting with $M=3$ and $N=6$, for an initial 
state made of a population-balanced half vortex with charges 
$(q,q')=(1,0)$. For $U=0.1\,J_\perp$ (resp. for $U=0$), the eigenvalues of the one-body density matrix are almost (resp. exactly) doubly degenerate, showing overlapped curves during the time interval exhibited in the figure, that is: $p_{1}\simeq p_2$, and $p_{3}\simeq p_4$.  
}
\label{vortex-groundB}
\end{figure}
In Fig.~\ref{vortex-groundB} we show the numerical results for 
an initial half-vortex state with $(q,q')=(1,0)$ and different values 
of the interaction $U/J=0.1,\,100$. The non-interacting dynamics 
(red lines) is also shown for comparison. Contrary to configuration A, 
this case only presents phase imbalance and not population imbalance between 
the rings, which according to the Josephson equations should translate 
into a particle current.
However, one can see from Fig.~\ref{vortex-groundB}, that it does 
not break the initial balance of population: $z(t) \simeq 0$ during 
the time evolution. Since the angular momentum carried by the 
tunneling particles is different in the top-to-bottom-ring current 
from the bottom-to-top-ring current, a non  vanishing angular 
momentum difference, or chiral current, is transferred between the 
rings, as it is shown in the top panel of Fig.~\ref{vortex-groundB}.

In spite of the differences in the initial state, the outcome 
is qualitatively similar to the evolution already shown for 
configuration A, and the coherent phase transfer between rings 
(if present at early stages) is eventually suppressed due to the 
presence of interparticle interactions. 
For small interactions $U/J_\perp=0.1$ (thick blue lines in 
Fig.~\ref{vortex-groundB}) the coherent behavior is preserved for 
a few Rabi cycles, and also presents revivals 
(not shown in the figure) for longer times. Simultaneously, 
$L_{\rm chi}(t)$ separates from the non-interacting result. Conversely, at 
large interaction values (symbols) the system becomes fragmented at 
the very beginning of the dynamics, and tends to a six-fragmented 
state. In general, the direct comparison between 
Figs.~\ref{vortex-groundA_1} and \ref{vortex-groundB} shows small 
differences in the chiral current or in the transition probability of
configurations A and B. A distinctive feature, however, can be observed
in the evolution of the population imbalance, whose uncertainty in configuration
B, around essentially zero average imbalance [see Fig. \ref{vortex-groundB}(b)],
presents a small variation even during the collapse of the coherent oscillations.

We have checked that an increase in the number of particles stretches 
the duration of the coherent-dynamics stage in the Josephson regime. In 
this way, our results tend to previous results obtained 
within the mean-field framework~\cite{gallemi3}, where long-duration 
coherent oscillations of half vortices were 
demonstrated in a two-component spinor condensate.

\begin{figure}[t]
\includegraphics[width=1\linewidth]{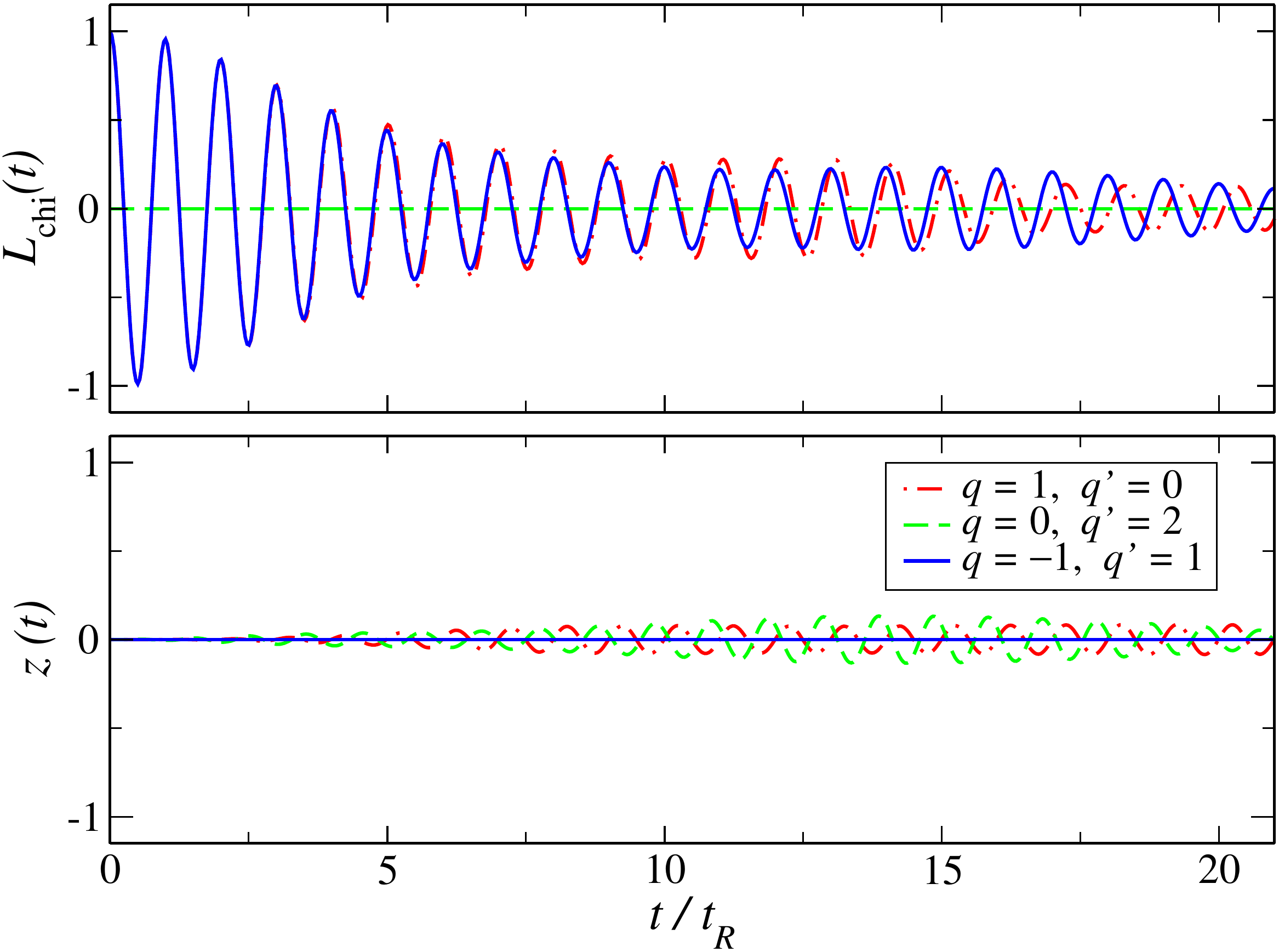} 
\caption{Dynamical evolution of the chiral current $L_{\rm chi}(t)$ 
and the population imbalance $z(t)$ for different fractional-vortex 
charges $(q,q')=(1,0)$ (red dashed-dotted lines), $(0,2)$(green dashed lines) 
and $(1,-1)$ (blue solid lines), and parameters 
$J/J_{\perp}=1$, $N=6$, $M=4$ and $U/J_{\perp}=0.1$. 
\label{diferent-quantizations}}
\end{figure}
Finally, we compare the fractional-vortex dynamics for several 
initial charges $(q,q')$. In Fig.~\ref{diferent-quantizations} we 
show the chiral current and the population imbalance in a system 
with $M=4$, $N=6$, $J/J_\perp=1$, and $U/J_\perp=0.1$, as a 
function of time, for $(q,q')=(1,0), \,(0,2),\, (1,-1)$. As in the 
previous cases, the evolution of the dynamical properties is 
sinusoidal with period $t_R=2\pi\hbar/2J_\perp$.
For $(1,0)$ and $(1,-1)$, the amplitude of the oscillations in the 
chiral current is smoothly damped, and only in the first Rabi cycles 
there is a quasi-complete exchange of the two initial vortex 
charges between the two rings. As expected for $(0,2)$, made of vortex states 
in a single ring situated at the center $q=0$ and  at the edge $q'=M/2$ of the 
Brillouin zone, there is no current and $L_{\rm chi}(t)=0$.

Figure \ref{diferent-quantizations} shows interesting differences 
in the population imbalance between $(1,-1)$, where it remains zero 
during the whole evolution, and the initial half-vortex states  
$(1,0)$ and $(0,2)$, where the population shows an oscillatory 
imbalance with varying amplitude. From general features of 
Josephson junctions, we attribute this different behavior to the 
differences in energy associated to the vortex charges involved in 
each configuration, since the Josephson equations predict that an 
energy variation across the junction modulates the particle current. 
Such modulation is absent in the $(1,-1)$ state, where the half vortices 
with charge $1$ and $-1$ are energetically degenerate.

\section{Conclusions and discussion}
\label{sec4}
We have studied the quantum tunneling dynamics of many-body vortices in
linearly coupled, discrete circuits. In a double ring geometry, we 
have considered population-imbalanced vortices and population-balanced, 
fractional vortices that can be first obtained in a stationary configuration of decoupled rings, and later monitored through the
observation of the population imbalance and the chiral particle flux during a coupled-ring dynamics.

The system preparation is well within the reach of current experimental techniques.
Protocols for the generation of ring-ladder lattices and the readout of flux qubits have been given in the literature (see e.g.~\cite{amico3}). 
The atoms could be loaded in two rings living in consecutive wells of
a perpendicular, tilted optical lattice, whose depth controlled the on-off switch of inter-ring 
tunneling  $J_\perp$, while the tilting provided the required initial population imbalance. By means of two-photon Raman processes, angular momentum could be imprinted on 
the atoms, and matter wave interferometry after time of flight expansion
can be used to observe the system phase patterns~\cite{Brachmann2011}. Alternatively, by using
laser-assisted tunneling in the ring lattice, a spatially  dependent  complex  tunneling  could be
imprinted to induce controlled vortex currents~\cite{Aidelsburger2013,Miyake2013}.

Our results show that, after preparing the initial vortices
at low interaction values, the subsequent dynamics is determined by the coherent 
oscillations of the initial vortex phase between the 
two rings, in a practically independent way of the initial vortex 
configuration. The vortex-flux connects current states with chiral symmetry,
and dually follows the usual sinusoidal particle current of 
the Josephson effect. The duration of the coherent regime increases with the 
number of particles, which makes the system more feasible for experimental 
realization. The high interaction regime, however, suppresses the 
superfluid tunneling dynamics through population self-trapping.

It is worth mentioning that the vortex tunneling is also sensitive 
to the number of couplings between rings. We have checked that when 
there is only one such coupling, like in the experiment of 
Ref.~\cite{Sturm2017}, tunneling processes are drastically reduced. 
Nevertheless, 
the study of the vortex tunneling dynamics as a function of the number of 
couplings between the two rings is beyond the scope of this work and will be 
addressed elsewhere.

\begin{acknowledgments}
We thank Alessio Celi for useful discussions. A. M. M. thanks 
GSCAEP at Beijing, where part of this work has been done, and 
especially to Xiaoquan Yu, for their hospitality. We acknowledge 
financial support from the 
Spanish MINECO and Fondo Europeo de Desarrollo Regional (FEDER, EU) 
under Grants No. FIS2017-87801-P and FIS2017-87534-P, 
and from Generalitat de Catalunya Grant No.  2017SGR533.  
A.E. is supported by Spanish MECD fellowship FPU15/03583.
\end{acknowledgments}

\appendix

\bibliography{refs.bib}
\end{document}